\documentstyle[aps,prb,epsf,floats,twocolumn]{revtex}

\newcommand{\tr}{\mbox{tr}\, }

\begin{document}
\draft
\title{Pumped current and voltage for an adiabatic quantum pump}

\author{M. L. Polianski and P. W. Brouwer}
\address{Laboratory of Atomic and Solid State Physics,
Cornell University, Ithaca, NY 14853-2501\\
{\rm \today}
\medskip \\ \parbox{14cm}{\rm
We consider adiabatic pumping of electrons through a quantum dot. 
There are two ways to operate the pump: to create a dc 
current ${\bar I}$ or to create a dc voltage ${\bar V}$.
We demonstrate that, for very slow pumping, ${\bar I}$ and ${\bar V}$ 
are not simply 
related via the dc conductance $G$ as $\bar I = \bar V G$. 
For the case of a 
chaotic quantum dot, we consider the statistical 
distribution of ${\bar V} G - {\bar I}$. Results are 
presented for the limiting cases of a dot with 
single channel and with multichannel point contacts. 
\smallskip
\pacs{PACS numbers: 73.23.-b, 72.10.Bg}}}
\maketitle

\section{Introduction}\label{intro}
An electron pump is a device that converts a periodic variation
of its characteristics into a time-independent electric current.\cite{thouless}
Such characteristics can be ``macroscopic'', like the charge 
on the device or the conductance of point contacts, or
``microscopic'', such as the location of a scatterer or the
magnetic flux threading the sample.
When there are two characteristics of the device that can be
varied harmonically with a frequency $\omega$ and phase difference 
$\phi$, pumping
of electrons already occurs in the adiabatic limit $\omega \to 0$.
In that case the pumped current is proportional to $\omega \sin \phi$, 
and changes sign when the phase relationship between the parameters
is reversed. Adiabatic electron pumps have been realized
experimentally in (arrays of) Coulomb blockaded quantum dots, using 
the voltages
on plunger gates, and/or the transparencies of the contacts as pumping
parameters.\cite{Kouwenhoven,Pothier,Oosterkamp}

In this paper, we consider an adiabatic electron pump that
consist of a semiconductor 
quantum dot coupled to two electron reservoirs
by means of ballistic point contacts.\cite{exp} 
Variation of two gate voltages allow 
for small changes of the shape of the dot, and thus for the
flow of a dc\ current. Following a proposal by Spivak 
{\em et al.},\cite{spivak}
such a device has been built and investigated by Switkes 
{\em et al.}\cite{switkes} 
The device of Ref.\ \onlinecite{switkes} is referred to
as an ``adiabatic quantum pump'', because
the variation of the gate voltages predominantly affects the 
quantum interference of the electrons in the quantum dot, not 
their classical trajectories. 
An important property of the quantum dot used in the experiment
of Ref.\ \onlinecite{switkes} is that its classical dynamics is
chaotic. As a result, the magnitude and the sign of the expected
pumped current $\bar I$ are subject to mesoscopic fluctuations. Since 
these fluctuations are
large, the mean $\langle \bar I \rangle$ and variance $\langle \bar
I^2 \rangle$ are insufficient to describe the ensemble,
and one needs to know the entire probability distribution 
$P(\bar I)$.

Theoretical analysis
has focused on the dc\ current $\bar I$
pumped through the dot,\cite{tok,zhou,shutenko,andreev,symmetry,Avron}  
\begin{equation}
  \bar I = {1 \over T} \int_0^{T} dt I(t), \label{eq:dcI}
\end{equation}
$T = 2\pi/\omega$ 
being the period of the pumping cycle. However, experimentally,
the
preferred measurement is that of the dc\ voltage $\bar V$ that the 
electron pump generates,\cite{switkes}
\begin{equation}
  \bar V = {1 \over T} \int_0^{T} dt V(t). \label{eq:dcV}
\end{equation} 
Naively, one might expect that, for small pumping amplitudes,
$\bar I$ and $\bar V$ are related via
the dot's conductance $G$ as $\bar I = \bar V G$. However, as we
show in this paper, this naive ``Ohm's law'' is not always true,
depending on whether the pumping frequency $\omega$
is small or large compared
to the charge-relaxation rate $\gamma$ of the reservoirs. (For
adiabatic pumping, $\omega$ must always be small compared to the
charge relaxation rate of the quantum dot.) For $\omega \ll \gamma$
and if the number $N$ of propagating 
channels in the point contacts between the quantum dot and the electron
reservoirs is small, the pumped current $\bar I$ and the difference
$\bar D = \bar V G - \bar I$ can actually be of comparable
magnitude.

A qualitative explanation why the pumped voltage and current are
not related via the simple relation $\bar V G - \bar I$ 
in the limit $\omega \ll \gamma$ follows from
the observation that both the pumped current and the
pumped voltage have dc\ and ac\ components. 
For slow pumping, the electron pump generates a bias voltage
that counteracts both the dc\ and ac\ currents generated in the
dot.
Since the conductance
$G$ itself also varies in time, $V_{\rm ac}G$ has a
 dc\ component. It is this additional
rectified dc\ component of the current
that is responsible for the difference between 
$\bar V G$ and $\bar I$ for an adiabatic electron pump.
When pumping is faster than the charge relaxation 
rate of the reservoirs, no ac bias voltage is generated to balance
the ac\ current, and 
the difference between $\bar V G$ and $\bar I$ disappears.

The purpose of this paper is to find the distribution of the
difference $\bar V G - \bar I$
between pumped current and pumped voltage for adiabatic
pumping of electrons through a chaotic quantum dot and to compare 
it to the distributions of $\bar I$ and $\bar V$. For a chaotic
dot, these distributions have a universal form, independent of
details of the pumping mechanism or the shape of the quantum
dot. In section \ref{model}
we use the scattering approach to present a quantitative theory
for the pumped voltage $\bar V$, and the difference 
$\bar D = \bar V G - \bar I$.  
In section \ref{dbar} we then evaluate the distribution 
of $\bar V G - \bar I$ for 
an ensemble of chaotic quantum dots, using random matrix theory.
We conclude in Sec.\ \ref{conclusion}.
\begin{figure}
\epsfxsize=0.8\hsize
\hspace{0.05\hsize}
\epsffile{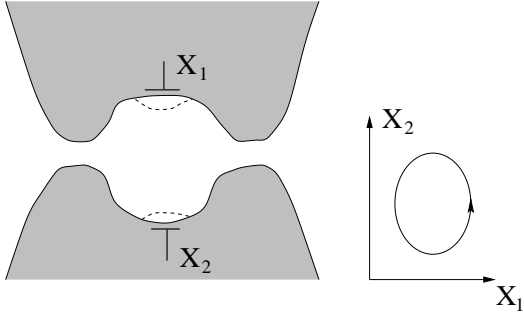}
\vglue +0.2cm
\caption{
\label{fig:1}
Schematic of a chaotic quantum dot whose shape can be changed by 
varying gate voltages $X_1$ and $X_2$ (left). In one cycle $X_1$ and 
$X_2$ trace out a contour in $(X_1,X_2)$ space (right).}
\end{figure}


\section{Pumped current and voltage}\label{model}
We consider a quantum dot coupled to two electron reservoirs 
via point contacts, see Fig.\ \ref{fig:1}. The shape of the
quantum dot is varied periodically by variation of two gate 
voltages, represented by dimensionless parameters $X_1$ and $X_2$.
Alternatively, $X_1$ or $X_2$ can represent the value of
an applied magnetic field, or any other parameter that
characterizes the quantum dot.

\begin{figure}
\epsfxsize=0.85\hsize
\hspace{0.0\hsize}
\epsffile{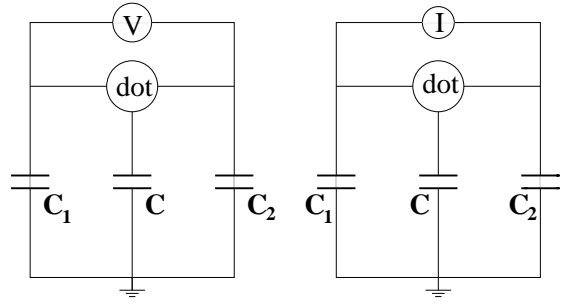}
\vglue +0.2cm
\caption{
\label{fig:2}
Equivalent circuit for a measurement of the pumped voltage through the
 dot (left) and of the pumped current (right). $C_i$ and $C$ are 
geometrical capacitances of the $i-$th reservoir and the dot.}
\end{figure}

As discussed in the
introduction, the electron pump can be characterized
experimentally via a direct measurement of the pumped current, 
or via a measurement of the voltage $V(t)$ between
the two reservoirs generated as a result of the pumping of
charge through the quantum dot. A formula for the dc\ current 
has been derived in Refs.\ \onlinecite{tok,zhou}.
To derive a formula for the dc\ voltage $\bar V$, we introduce 
a simple model for the 
quantum dot and the two electron reservoirs, see Fig.\
\ref{fig:2}. The dot and the reservoirs $1$, $2$
are connected to
a screening gate via capacitances $C$ and $C_{1}$, $C_{2}$.
Following Refs.\ \onlinecite{buttiker1,buttiker2}, 
we introduce the emissivity $e\delta
q(m)/\delta X_j$, which is the charge that exits the dot
through point contact $m$ ($m=1,2$) when the parameter $X_j$ ($j=1,2$)
is changed adiabatically by an amount $\delta X_j$. Then the
total current flowing through contacts $1$ and $2$ reads
\begin{mathletters}\label{general}
\begin{eqnarray}
  I_1(t) &=& e\sum_{i=1}^2 \frac{\delta q(1)}{
\delta X_i}\frac{dX_i}{dt}+[V_1(t) - V_2(t)] G, 
  \label{eq:general1}\\
  I_2(t) &=& e\sum_{i=1}^2 \frac{\delta q(2)}{
\delta X_i}\frac{dX_i}{dt}+[V_2(t) - V_1(t)] G,
  \label{eq:general2}
\end{eqnarray}
\end{mathletters}%
where $G$ is the dc\ cconductance of the quantum dot. Here we
assume that all variations are made slowly on the scale of the
dwell time of the quantum dot.

When current is measured, the voltages of the two reservoirs
are equal, $V_1(t) = V_2(t)$. Hence the dc\
current $\bar I$ can be found from integration of Eq.\ 
(\ref{eq:general1})
or (\ref{eq:general2}).\cite{tok} 
Using Stokes' theorem, $\bar I$ can be
rewritten as an integral over the surface area $S$ enclosed by the
contour of the parameters $X_1$ and $X_2$ in the ($X_1$, $X_2$)
plane,
\begin{mathletters}
\begin{eqnarray}\label{eq:resIint}
\bar I &=& {e \omega \over 2 \pi} \int dX_1dX_2 \bar i(X_1,X_2),
\end{eqnarray}
where
\begin{eqnarray}
\bar i &=& \frac{\partial}{\partial X_2}
\frac{\partial q(1)}{\partial X_1}-\frac{\partial}{\partial X_1}
\frac{\partial q(1)}{\partial X_2} \nonumber \\
&=& - \frac{\partial}{\partial X_2}
\frac{\partial q(2)}{\partial X_1}+\frac{\partial}{\partial X_1}
\frac{\partial q(2)}{\partial X_2}.
\label{eq:resi}
\end{eqnarray}
\end{mathletters}%
When voltage is measured, the result depends
on whether variation of the parameters
$X_1$ and $X_2$ is fast or slow compared to the charge relaxation
rates $\gamma_{1,2} \sim G/C_{1,2}$ of the reservoirs. 
If the variation is fast compared to $\gamma_{1,2}$ (but still
slow compared to the charge relaxation rate of the quantum dot), 
the voltage difference $V_{1} - V_{2}$ is essentially time 
independent and takes the value
\begin{eqnarray}
  \bar V = \bar I/\bar G,
\end{eqnarray}
where $\bar G$ is the conductance averaged over one cycle,
\begin{eqnarray}
  \bar G = {1 \over T} \int_0^{T} dt G(X_1(t),X_2(t)).
\end{eqnarray}
In the opposite limit $\omega \ll \gamma_{1,2}$,
which is the case we'll consider in the remainder of the paper, 
one has $I_1(t) = \eta I_2(t)$, where $\eta = C_{1}/C_{2}$ is 
a numerical
coefficient describing the capacitive division between the
two reservoirs. Combining this with Eq.\ (\ref{general}), we find
\begin{eqnarray}
V_1(t) - V_2(t) &=&\frac{e}{G(1+\eta)}\sum_{i=1}^{2}\left(\frac{\delta q(1)}{
  \delta X_i}\frac{dX_i}{dt}-\eta\frac{\delta q(2)}{\delta X_i}
\frac{dX_i}{dt} \right)\nonumber\\
&&\mbox{}
\end{eqnarray}

The dc\ voltage $\bar V$ is then found by integration over the
period $T$, with the result
\begin{mathletters}
\begin{eqnarray}\label{eq:resVint}
\bar V&=&\frac{h \omega}{4 \pi e}\int dX_1dX_2 \bar v(X_1,X_2), \\
\label{eq:resv}
\bar v &=& {1 \over 1 + \eta} \left[\frac{\partial}{\partial X_2}(\frac 1g
\frac{\partial q(1)}{\partial X_1})-\frac{\partial}{\partial X_1}
(\frac 1g\frac{\partial q(1)}{\partial X_2})\right]
  \nonumber \\ && \mbox{} - {\eta \over 1 + \eta} 
\left[\frac{\partial}{\partial X_2}(\frac 1g
\frac{\partial q(2)}{\partial X_1})-\frac{\partial}{\partial X_1}
(\frac 1g\frac{\partial q(2)}{\partial X_2})\right],
\end{eqnarray}
\end{mathletters}%
where $g =hG/2e^2$ is the dimensionless conductance. 
For small harmonic variations of the parameters,
$X_1(t) = \delta X_1 \cos (\omega t)$ and 
$X_2(t) = \delta X_2 \cos(\omega t + \phi)$, the integrations
are trivial, and one finds
\begin{eqnarray}\label{res2}
  \bar I &=& {1 \over 2} e \omega \delta X_1 \delta X_2 \bar i 
\sin \phi
  \label{eq:resIinf}
\end{eqnarray}
for the pumped current, and
\begin{eqnarray}
\bar V &=& \frac{h \omega}{4 e} \delta X_1 \delta X_2 \bar v \sin \phi 
  \label{eq:resVinf}
\end{eqnarray}
for the pumped voltage.

If the conductance $G$ were constant in a cycle, i.e., $G$
would not depend on $X_1$ and $X_2$, Eqs.\ 
(\ref{eq:resi}) and (\ref{eq:resv})--(\ref{eq:resVinf}) 
give the identity $\bar V G=\bar I$:
the pumped current measured at zero bias and the pumped voltage
measured at zero current are related by Ohm's law. 
However, in general 
$G$ does depend on $X_1$ and $X_2$, and this ``Ohm's law'' does 
not hold. The deviation is described by the difference
\begin{equation}
  \bar D = \bar V G - \bar I.
\end{equation}
For small $\delta X_{1}$ and $\delta X_2$, we find from
Eqs.\ (\ref{eq:resi}), (\ref{eq:resv}),
\begin{mathletters}
\begin{eqnarray}
  \bar D &=& {1 \over 2} e\omega \delta X_1 \delta X_2 \bar d \sin \phi, \\
  \bar d &=&
  \label{vg}
  {1 \over 1 + \eta}
\left[\frac 1g\frac{\partial g}{\partial X_1}\frac{\partial q(1)}
{\partial X_2}-\frac 1g\frac{\partial g}{\partial X_2}\frac{\partial q(1)}
{\partial X_1}\right] \nonumber \\ && \mbox{} - {\eta \over 1 + \eta}
  \left[\frac 1g\frac{\partial g}{\partial X_1}\frac{\partial q(2)}
{\partial X_2}-\frac 1g\frac{\partial g}{\partial X_2}\frac{\partial q(2)}
{\partial X_1}\right].
\end{eqnarray}
\end{mathletters}%

The derivatives to $X_1$ and $X_2$ appearing in the above formulae
are derivatives taken at constant values of the electro-chemical
potential $\mu$ of the reservoirs. These are not
necessarily equal to derivatives taken at
a constant value of the (self-consistent\cite{abg}) electrostatic
potential $V_{\rm sc}$ inside the quantum dot. 
Changes of $V_{\rm sc}$
occur in a pumping cycle, because the 
total charge on the dot may vary during the
pumping cycle. For technical reasons, it is preferred to treat
$V_{\rm sc}$, or, equivalently, the kinetic energy $E = \mu - 
V_{\rm sc}$, as an independent parameter, and to take derivatives
at constant $E$. The above equations 
for pumped voltage and current can be rewritten using 
derivatives at constant $E$ if we substitute the parametric 
derivatives $\partial /\partial X_1$ and $\partial/\partial X_2$ 
in Eqs.\ (\ref{eq:resi}), (\ref{eq:resv}), and (\ref{vg}) 
by\cite{buttiker1,buttiker2}
\begin{mathletters}
\begin{eqnarray}\label{eq:vgandc}
\left. \frac{\partial}{\partial X} \right|_{\mu}
 &\to & \left. \frac{\partial}{\partial X} \right|_{E}+
 \frac{\partial E}{\partial X}\frac{\partial}{\partial E},\\
\frac{\partial E}{\partial X_i}&=&-\frac{\partial q/\partial X_i}
{C/2 e^2 +\partial q/\partial E}.
\end{eqnarray}
\end{mathletters}%
Here $C$ is the capacitance of the quantum dot. In Eq.\ 
(\ref{eq:vgandc}), we abbreviated
\begin{equation}
  {\partial q \over \partial X_j} =
  {\partial q(1) \over \partial X_j}
  + {\partial q(2) \over \partial X_j}.
\end{equation}
For a realistic quantum dot, the charging energy $e^2/C$ is much
larger than the mean level spacing $\Delta$. In that limit, one 
finds that the dot charge remains constant during the pumping
cycle, $I_1(t) = -I_2(t)$ for all time. As a consequence, the
pumped voltage $\bar V$ and the difference $\bar D = \bar V G -
\bar I$ lose their dependence on the capacitive division $\eta$.

In the absence of inelastic processes and for low temperatures
(temperature $T$ below the mean level spacing $\Delta$ in the
quantum dot), the emissivities $\partial
q(m)/\partial X_j$ and the conductance $G$ can be expressed in terms
of the scattering matrix $S$ of the quantum dot and its derivatives
to $X_1$, $X_2$, and $E$. The matrix $S$ has dimension $2N$, where
$N$ is the number of propagating channels in each point contact;
it is unitary (unitary symmetric) in the presence (absence) of a
time-reversal symmetry breaking magnetic field. The derivatives
of $S$ are parameterized via 
hermitian matrices $R$, $R_j$, defined as\cite{Smith}
$$
R=-i{\Delta \over 2 \pi}
  \frac{\partial S}{\partial E}S^\dagger,\ R_j=-i\frac{\partial S}
{\partial X_j}S^\dagger.
$$
Then the emissivities $\partial q(m)/\partial E,\ \partial q(m)/
\partial X$, $m=1,2$, are given by\cite{buttiker2}
\begin{mathletters}
\begin{eqnarray}
{\partial q(m) \over \partial E}  &=&  \frac{1}{\Delta}\mbox{Re tr}\,
  P_m R,\\
  {\partial q(m) \over \partial X_j} &=& \frac{1}{2\pi}\mbox{Re tr}\,
  P_m R_j.
\end{eqnarray}
\end{mathletters}%
Here $P_{1} = 1 - P_{2}$ is a diagonal matrix with elements
$(P_{1})_{jj} = 1$ if $j \le m$ and zero otherwise.
The dc\ conductance $g$ is given by the Landauer formula,
\begin{eqnarray}
 g = \mbox{tr\,} S^\dagger P_1SP_2.
\end{eqnarray}

In the next section, we shall study the distribution of the 
dimensionless difference $\bar d =
\bar v g - \bar i$ for the case of a chaotic quantum dot.
\section{Distribution of $\bar D = \bar V G - \bar I$}\label{dbar}

For an ensemble of chaotic quantum dots, the statistical distribution
of the
scattering matrix and its derivatives is known from the 
literature.\cite{waves}
It takes its simplest form when the derative of $S$ is not
parameterized
by the matrices $R_{j}$, $R$, but by the symmetrized derivatives
$Q$, $Q_j$,
$$
 Q=S^{-1/2}RS^{1/2},\ Q_j=S^{-1/2}R_jS^{1/2}.
$$
In the prensence of time-reversal symmetry (labeled by the Dyson 
parameter $\beta=1$), the matrices $Q$, $Q_1$, and $Q_2$ are
real symmetric. When time-reversal symmetry is broken by a magnetic
field ($\beta=2$), $Q$, and $Q_1$, and $Q_2$ are hermitian. 
For ideal contacts, the joint distribution $P(S,Q,Q_1,Q_2)$
reads\cite{waves}
\begin{eqnarray}\label{p}
P&\propto&\left(1 + {2 e^2 \over C\Delta} \mbox{tr}\, Q
\right)  (\det Q)^{-N/2 - 2 (\beta N+2-\beta)}
  \nonumber \\ && \mbox{} \times
\exp\left[-\frac \beta 2\mbox{tr}\, \left(
Q^{-1}\right)\right] \Theta(Q) \\&& \nonumber \mbox{}
\times\exp\left[-\frac{\beta}{16}
\mbox{tr}\, \left(\left(Q^{-1}Q_1\right)^2 +
\left(Q^{-1}Q_2\right)^2
  \right)\right],
\end{eqnarray}
where $\Theta(Q)=1$ if all eigenvalues of $Q$ are positive and
$\Theta(Q)=0$ otherwise.
The capacitance $C$ appears in the distribution because the ensemble is
obtained by sweeping an external gate voltage, not the self-consistent
energy $E$.\cite{parametric}

To find the distribution $P(\bar d)$, we
first integrate over $Q_1$ and $Q_2$ at fixed
$S$ and $Q$, and then over $S$ and $Q$. The first integration
can be done analytically, since, for fixed $Q$,
$Q_1$ and $Q_2$ are Gaussian random matrices, see
Eq.\ (\ref{p}). The result of this
integration takes a simple form,
\begin{eqnarray}
  P(\bar d) &=& \left\langle {1 \over 2 \sigma} e^{-|\bar d|/\sigma}
  \right\rangle_{S,Q}, \label{eq:Pd}
\end{eqnarray}
where the brackets indicate the average over $S$ and $Q$ that
remains to be done. In equation (\ref{eq:Pd}) $\sigma$ is
positive function of $S$ and $Q$, given by
\begin{eqnarray}\label{sigmas}
  \sigma^2 &=&\left(\frac{16}{\beta g}\right)^2\left(
  [\tr A^2+ {\delta_{\beta,1} \over 2}
  \tr(PRSPR^TS^\dagger-\right.\nonumber
\\ &&\left. \vphantom{\delta_{\beta,1} \over 2}
  \mbox{} - PRPR)]\ \tr B^2 - (\tr A B)^2\right), 
\end{eqnarray}
where we abbreviated
\begin{eqnarray}
  \nonumber A &=& R \left( P - { \tr P R \over C\Delta/2e^2 + \tr R}
  \right),\\ 
\nonumber  B &=& R \left( \Lambda - 
  { \tr \Lambda R \over  C\Delta/2e^2 + \tr R}
  \right),
\end{eqnarray}
and $P = (P_1 - \eta P_2)/(1+\eta)$, $\Lambda =
i(P_1SP_2S^\dagger- S P_2 S^{\dagger} P_1)$.

For the remaining integrations over $S$ and $Q$ we consider
two limiting cases: multichannel point contacts ($N \gg 1$) and
single channel point contacts ($N=1$).

\subsection{Single-channel contacts}\label{1}

\begin{figure}
\vglue -0.45cm
\epsfxsize=0.9\hsize
\hspace{0.0\hsize}
\epsffile{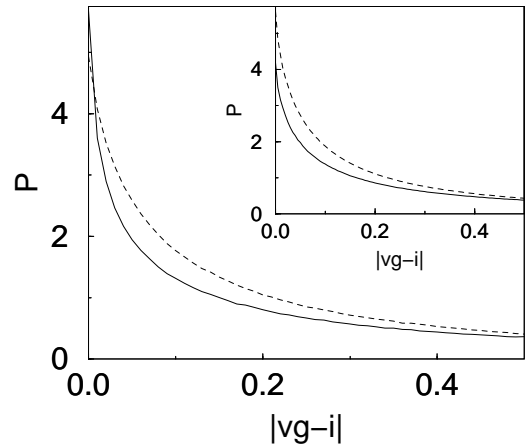}
\vglue +0.2cm
\caption{
\label{picvg}
Distribution of the normalized
difference $\bar d = \bar v g - \bar i$ between pumped voltage
$\bar v$ and pumped current $\bar i$ for single-channel point 
contacts, for the physically relevant limit $C \Delta \ll e^2$
(main figure). The opposite limit $C \Delta \gg e^2$ is shown in 
the inset for the case $\eta=0$.
Presence (absence) of time-reversal symmetry $\beta=1$ ($\beta=2$) 
is shown solid (dashed).}
\end{figure}

For $N=1$ the remaining number of variables is small, and can
be integrated over numerically, using the distribution (\ref{p}).
Results for the (physically
relevant) limit $C \Delta \ll e^2$ are shown in Fig.\
\ref{picvg}. In this limit, the distribution $P(\bar d)$ does
not depend on the capacitive division $\eta$ between the reservoirs.
The results for larger values of $C$ are not very different from
those shown in Fig.\ \ref{picvg}. This is illustrated in the inset
of Fig.\ \ref{picvg}, where we have shown the distribution
$P(\bar d)$ for the case of large capacitance $C \Delta \gg e^2$
and asymmetric reservoirs (capacitive division $\eta=0$).
For $\bar d$ close to zero, the distribution shows a cusp with 
a logarithmically divergent derivative at $\bar d=0$.
For $\bar d \gg 1$, the distribution $P(\bar d)$ has power-law
tails. For $C \Delta \ll e^2$ these are\cite{foottails} 
\begin{eqnarray}\label{result11}
P(\bar d) \propto \left\{ \begin{array}{ll}\bar d^{-3}, 
& \beta=1, \\
 \bar d^{-3}\log \bar d, & \beta=2. \end{array}\right.
\end{eqnarray}
The tails of the distribution
correspond to samples with an anomalously
large eigenvalue of $Q$, corresponding to an anomalously large 
dwell time $\tau_D$:\cite{Smith} a value of $\bar d$
in the tail of the contribution typically corresponds to a dwell time
$\tau_D \sim \tau_H \bar d^{2/\beta}$, $\tau_H = \hbar/\Delta$ being 
the Heisenberg
time. Since configurations with anomalously large dwell times are
more sensitive to dephasing or thermal smearing, such perturbations
will truncate the tails for $\bar d \gtrsim (\tau_{\phi}/\tau_H)^{
\beta/2}$ or $\bar d \gtrsim (\tau_H T)^{-\beta/2}$.

\begin{figure}
\vglue -0.45cm
\epsfxsize=0.9\hsize
\hspace{0.0\hsize}
\epsffile{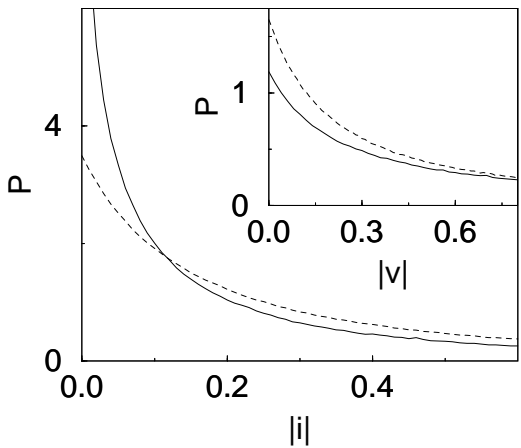}

\caption{
\label{piciv}
Distribution of normalized pumped current $\bar i$ 
and (inset) voltage $\bar v$ for single-channel 
contacts. Presence 
(absence) of time-reversal symmetry $\beta=1$ ($\beta=2$) is 
shown solid (dashed).}
\end{figure}

For comparison, we have also calculated the distribution of the 
pumped current $\bar i$ and voltage $\bar v$.
For single-channel contacts, the distribution again takes the form 
(\ref{eq:Pd}), with $\sigma$ replaced by $\sigma_i$ and $\sigma_v$,
respectively. Expressions for $\sigma_i$ and $\sigma_v$ can be found
in the appendix. 
The resulting distributions are shown in Fig.\ \ref{piciv}. 
(The distribution of the current was calculated previously in 
Ref.\ \onlinecite{tok}.)
The main conclusion upon comparison of Figs.\ \ref{picvg} and
\ref{piciv} is that, for
single-channel point contacts, the distributions of $\bar i$
and of $\bar d = \bar v g - \bar i$ have comparable widths.
Hence, for $N=1$, deviations from ``Ohm's law'', as characterized
by $\bar d = \bar v g - \bar i$ are of the same order as the
pumped current $\bar i$ itself.

\subsection{Multi-channel contacts}\label{2}
The distribution of $\bar d$ in the limit $N \gg 1$ can be directly
obtained from Eq.\ (\ref{eq:Pd}).
The integration over the unitary matrix $U$ of eigenvectors of $R$
and over the scattering matrix $S$ is performed using the method
of Ref.\ \onlinecite{unitary}. 
For the remaining integration over the eigenvalues $\tau_i,
i=1,\ldots,2N,$ of the matrix $R$ it is sufficient to know their
density,\cite{waves} 
\begin{eqnarray}
\rho(\tau)=
  \sum_{j=1}^{2N} \langle \delta(\tau_j-\tau) \rangle
  =  \frac{N}{\pi\tau^2}\sqrt{(\tau_{+}-\tau)(\tau-\tau_{-})},
\end{eqnarray}
where $\tau_{\pm}=({3\pm\sqrt{8}})/{2N}$.
This gives the result
\begin{equation}\label{multichannel}
\langle \sigma^2 \rangle
=\frac{4}{\beta N^4}\left[1+\frac 4\beta\left(\frac{1-\eta}
{1+\eta}\right)^2\left(\frac{\Delta}{\Delta+2e^2/C}\right)^2\right].
\end{equation}
Calculation of higher moments of $\sigma$ shows that fluctuations of 
$\sigma$ are small compared to the average as $N \to \infty$. Hence,
we can conclude that the distribution of $\bar d$ in multi-channel
limit is of the Poissonian form (\ref{eq:Pd}), with $\sigma^2$ 
given by Eq.\ (\ref{multichannel}).

This is to be contrasted with the distribution of the pumped current
$\bar i$, which is Gaussian for large $N$, 
with zero mean and with root mean 
square\cite{tok,shutenko}
\begin{equation}\label{isigma}
  \langle \bar i^2 \rangle^{1/2} = {1 \over \pi N}.
\end{equation}
Hence we conclude that, for large $N$, typically $\bar d$ is a factor
$\sim N$ smaller than the pumped current, and can be neglected in 
a measurement. Hence, in the limit $N\to\infty$ the 
expectation of ``Ohm's law'' $\bar v g=\bar i$ holds, 
and one readily concludes
that the pumped voltage $\bar v$ has a Gaussian distribution with
zero mean and with root mean square
\begin{eqnarray}\label{vsigma}
  \langle \bar v^2 \rangle^{1/2}
  &=& {2 \over \pi N^2}.
\end{eqnarray}

\section{conclusion} \label{conclusion}
In summary, for adiabatic pumping of electrons through 
a chaotic quantum dot, we have derived expressions of the pumped
current $\bar I$ (in case of a current measurement) or the pumped 
voltage $\bar V$
(in case of a voltage measurement) in terms of the scattering 
matrix of the quantum dot. Pumped current and voltage are not simply related
by the dot's conductance $G$. We have calculated the distribution
of the difference $\bar V G - \bar I$ for small pumping amplitudes,
which is universal for a chaotic quantum dot. If the number $N$ of 
propagating channels in the contacts
between the quantum dot and the reservoirs is one, $\bar I$ and
$\bar V G - \bar I$ can be of the same size; if $N \gg 1$, 
$\bar V G - \bar I$ is typically a factor $N$ smaller than
$\bar I$. Our results are valid in the limit of slow pumping,
where the pumping frequency $\omega$ is much smaller than the
charge relaxation rate $\gamma$ of the reservoirs. If $\omega \gg
\gamma$, the difference $\bar V G - \bar I$ is suppressed.

The results obtained here are important in view of the interpretation
of the experiment of Ref.\ \onlinecite{switkes}. The observations
of that experiment can also be explained if the observed dc voltage
is the result of rectification of ac displacement currents generated
by the time-dependent gate voltages that should drive the electron
pump.\cite{rectif} Therefore, it is important to identify signatures
that distinguish adiabatic pumping from mere
rectification of displacement currents. For the case of a current
measurement, two such signatures are the magnetic field symmetry 
and the typical size of the pumped
current.\cite{shutenko,rectif}
Except for the
case $N=1$, our results allow to translate these signatures to a 
voltage measurement as well.\cite{n=1} Further, the relation
between pumped voltage and pumped current provides a third signature
of a adiabatic pumping: For few-channel point contacts, $\bar I$,
$\bar V G$, and the difference $\bar V G - \bar I$ are all random 
and of comparable magnitude for a quantum pump, while, if the dc 
signal is due to rectification, there is a fixed relationship 
$\bar I \propto G^2 \bar V$, the proportionality constant being 
non-universal.\cite{rectif}

We thank C. Marcus for discussions. This work was supported by the
NSF under grant no.\ DMR-0086509 and by the Sloan foundation.

\appendix
\section{}

Below we present results for the ``intermediate''
distributions of
the normalized current $\bar i$, voltage $\bar v$, 
and of the difference $\bar d = \bar v g - \bar i$ after
integration over the matrices $Q_1$ and $Q_2$ at fixed $S$ and 
$Q$, but before integration over $S$ and $Q$,
for the case of a chaotic quantum dot with two single-channel
point contacts. For that case, the distribution
 of $\bar d$ is given by
\begin{eqnarray}
  P(\bar d) &=& \left\langle {1 \over 2 \sigma_d} e^{-|\bar d|/\sigma_d}
  \right\rangle_{S,Q}, \label{eq:PdApp} 
\end{eqnarray}
where the brackets $\langle \ldots \rangle_{S,Q}$ denote the remaining
average over the scattering matrix $S$ and the symmetrized
time-delay matrix $Q$. The distributions for $\bar i$, $\bar v$ have 
the same form with $\sigma_d$ replaced by $\sigma_i$ and $\sigma_v$,
respectively. However, we should note that, unlike for the
difference $\bar d$, the form (\ref{eq:PdApp}) does not hold for the 
distributions of $\bar i$ and $\bar v$ when $N>1$. 

Below we list (statistical) expressions for $\sigma_d$, $\sigma_i$
 and $\sigma_v$ for $N=1$ for the case $C \Delta \ll e^2$. 
We introduce the eigenvalues $\tau_1,\tau_2$ of the normalized
time-delay matrix $R$. Their distribution can be
found in Ref.\ \onlinecite{waves}. Further, we introduce two
independent random variables $t$ (uniformly distributed between $0$ and
$1$) and $\phi$ (uniformly distributed between $0$ and $2\pi$)
that arise from the randomly distributed eigenvectors of $R$ and 
the phases of the scattering matrix $S$. Finally, the equations
for $\sigma_d$, $\sigma_i$, and $\sigma_v$ contain the dimensionless 
conductance $g\in[0,1]$, which has distribution $P(g) = (\beta/2)
g^{-1 + \beta/2}$. We then find
\begin{eqnarray*}
\sigma_{d,\beta=1}^2&=&256\frac{(\tau_1\tau_2)^3}{(\tau_1+\tau_2)^2}
\frac{(1-g)^2}{g},\\
\sigma_{i,\beta=1}^2&=&256\frac{(\tau_1\tau_2)^3}{(\tau_1+\tau_2)^2}g,\\
\sigma_{v,\beta=1}^2&=&256\frac{(\tau_1\tau_2)^3}
{(\tau_1+\tau_2)^2}\frac{1}{g^3},\\
\sigma_{d,\beta=2}^2&=&\frac{1-g}{g}\left(\frac{8\tau_1\tau_2}
{\tau_1+\tau_2}\right)^2\\ &&\mbox{}\times\left(\tau_1\tau_2+
t(1-t)(\tau_1-\tau_2)^2\sin^2 \phi\right),\\
\sigma_{i,\beta=2}^2&=&(4\tau_1\tau_2)^2\left(1-4t(1-t)\left(
\frac{\tau_1-\tau_2}{\tau_1+\tau_2}\right)^2\right),\\
\sigma_{v,\beta=2}^2&=&\frac{1}{g^3}\left(\frac{8\tau_1\tau_2}
{\tau_1+\tau_2}\right)^2 \left[\tau_1\tau_2+(\tau_1-\tau_2)^2\right.
 \nonumber \\ && \left.\times\left(2\sqrt{(1-g)t(1-t)}\sin\phi+(1-2t)\sqrt{g} \right)^2\right].
\end{eqnarray*}



\begin{references}
\bibitem{n=1}
For single-channel contacts $N=1$, $\bar v$ is expressed in terms of 
the scattering matrix $S$ as 
\begin{eqnarray*}
\bar v &=& \frac{1}{(1+\eta)}\frac{2}{|S_{12}|^4}\mbox{Im }\left(\frac
{\partial S_{11}}{\partial X_1}\frac{\partial S_{11}^*}{\partial X_2}-
\eta\frac{\partial S_{22}}{\partial X_1}\frac{\partial S_{22}^*}
{\partial X_2}\right).
\end{eqnarray*} 
Since $S(B)=S^T(-B)$ under reversal of a magnetic field $B$, this 
equation results in $\bar v(-B)=\bar v(B)$. There is no such magnetic
field symmetry of the pumped voltage for $N > 1$.
\end{references}
\end{document}